\documentclass[conference]{IEEEtran}
\usepackage{subfig}
\usepackage{graphicx}
\usepackage{epsfig}
\usepackage{amssymb}
\usepackage{amsmath}
\usepackage{algorithm}
\usepackage[noend]{algpseudocode}
\usepackage{multirow,url}
\makeatletter
\def\BState{\State\hskip-\ALG@thistlm}
\makeatother

\begin{document}
\title{Community-based Outlier Detection \\for Edge-attributed Graphs}

\author{\IEEEauthorblockN{Supriya Pandhre}
\IEEEauthorblockA{Indian Institute of Technology, Hyderabad\\Hyderabad, India\\
cs15mtech11016@iith.ac.in}
\and
\IEEEauthorblockN{Manish Gupta}
\IEEEauthorblockA{Microsoft\\Hyderabad, India\\
gmanish@microsoft.com}
\and
\IEEEauthorblockN{Vineeth N Balasubramanian}
\IEEEauthorblockA{Indian Institute of Technology, Hyderabad\\Hyderabad, India\\
vineethnb@iith.ac.in}
}
\maketitle

\begin{abstract}
The study of networks has emerged in diverse disciplines as a means of analyzing complex relationship data. Beyond graph analysis tasks like graph query processing, link analysis, influence propagation, there has recently been some work in the area of outlier detection for information network data. Although various kinds of outliers have been studied for graph data, there is not much work on anomaly detection from edge-attributed graphs. In this paper, we introduce a method that detects novel outlier graph nodes by taking into account the node data and edge data simultaneously to detect anomalies. We model the problem as a community detection task, where outliers form a separate community. We propose a method that uses a probabilistic graph model (Hidden Markov Random Field) for joint modeling of nodes and edges in the network to compute \emph{Holistic Community Outliers (HCOutliers)}. Thus, our model presents a natural setting for heterogeneous graphs that have multiple edges/relationships between two nodes. EM (Expectation Maximization) is used to learn model parameters, and infer hidden community labels. Experimental results on synthetic datasets and the DBLP dataset show the effectiveness of our approach for finding novel outliers from networks.
\end{abstract}

% no keywords - Anomaly Detection, Graph with multiple edges, Probabilistic Graphical Model
%\IEEEpeerreviewmaketitle
\section{Introduction}
\label{sec:introduction}

%General discussion around applications of outlier detection, and outlier detection from networks
Outlier (or anomaly) detection is a very broad field which has been studied in the context of a large number of research areas like statistics, data mining, sensor networks, environmental science, distributed systems, spatio-temporal mining, etc. Many algorithms have been proposed for outlier detection in high-dimensional data, uncertain data, stream data and time series data. By its inherent nature, network data provides very different challenges that need to be addressed in a special way. Network data is gigantic, contains nodes of different types, rich nodes with associated attribute data, noisy attribute data, noisy link data, and is dynamically evolving in multiple ways. Outlier detection on networks for various applications can be useful for: (1) identification of interesting entities or sub-graphs, (2) data de-noising (both with respect to the network nodes and edges), (3) understanding the anomalous temporal behavior of entities, and (4) identification of new trends or suspicious activities in both static and dynamic scenarios. Some examples of network outliers include users who spread rumors on Twitter, unusual but successful associations of persons with various organizations in an organizational network, sensor with anomalous readings in a sensor network, anomalous traffic between two components in a distributed network, snapshot with broken correlation between network properties across time, etc.

%main outlier detection methods for static and dynamic outlier detection
Given a static graph, one can find node, edge or subgraph outliers. For dynamic graphs, the outliers could be time stamps where the properties of the snapshot are significantly different from the properties of other snapshots. In the dynamic scenario, one can also identify a node (or an edge, or a subgraph) which shows anomalous behavior across time as an outlier. Popular outlier detection methods for static graphs include the Minimum Description Length (MDL) method~\cite{chakrabarti2004_ecmlpkdd}, frequent subgraph mining, egonet analysis~\cite{akoglu2010_pakdd}, and community analysis~\cite{gao2010_kdd,gupta2013community}. Popular methods for outlier detection for dynamic graphs include graph similarity based methods~\cite{pincombe05_asorb}, temporal spectral analysis~\cite{ide04_kdd}, temporal structural connectivity analysis~\cite{aggarwal11_icde}, and temporal community analysis~\cite{gupta12_pkdd,gupta12_kdd}.

%what is the novel aspect of the proposed outlier detection method and how does it relate to past work

While most work in the past on network outlier detection has focused on graphs with node data and links, there is very little work in the area of edge-attributed graphs. In some cases, node data and linkage could be normal even when edge data is abnormal, in the context of nodes on which the edge is incident. With the rich variety of interactions in various complex graphs across a variety of domains, it is critical to incorporate such edge attributes in finding novel outliers. If we could assign a latent community to every node and every edge, a HCOutlier node is one which is linked to many nodes of another community, or has incident edges belonging to another community. 

The following real-world examples demonstrate the importance of HCOutliers.

%poster child example to explain the novelty of new outlier detection and also explain its importance in a domain
\noindent{\underline{Organizational Graph Example}}: A graph of people working in a company is called an organization graph. Edges represent frequent communication. Role of a person can be considered as the community label. Usually a chemist in a company would communicate frequently with other chemists. But a suspicious HCOutlier node would not just communicate frequently with members of other roles, but also the community of the communications would be different from the one expected for its role. Such a person could be an activist planning a malicious activity with other folks in the company, or a star performer working on a secret collaborative project.

\noindent{\underline{Co-authorship Graph Example}}: A co-authorship graph contains authors as nodes. Two authors are connected if they have published a paper together. On such graphs, research areas can be considered as latent communities. Most authors collaborate with other authors, and publish papers \emph{in the same research area}. An HCOutlier would be an author who: (1) collaborates frequently with authors from other research areas, and (2) collaborates with other authors of same or different community on papers which belong to other research areas. Such authors perform inter-disciplinary research and are vital in cross-pollination of concepts across research areas.

\noindent{\underline{Co-actorship Graph Example}}: A co-actorship graph contains actors as nodes. Two actors are connected if they have worked in a movie together. On such graphs, movie genres can be considered as latent communities. Most actors collaborate with other actors, and work in movies \emph{in the same genre}. An HCOutlier would be an actor who: (1) collaborates frequently with actors from other genres, and (2) collaborates with other actors of same or different genre in movies which belong to other genres. HCOutlier actors are ones who have built a reputation for being multi-faceted actors with an eclectic range of movies.

\noindent{\underline{Limitations of Past Approaches}}: 
In the past, various approaches have been proposed by looking only at node data and ignoring network structure. Clearly, HCOutliers cannot be detected using such a global approach because HCOutlier nodes belong to popular communities. There has been work on finding community outliers for both homogeneous~\cite{gao2010_kdd} and heterogeneous graphs~\cite{gupta2013community} by considering node data and link structure but ignoring edge data. Clearly, such an approach can identify some HCOutliers who are linked to nodes with a different community label. But they cannot identify a HCOutlier who could be connected to nodes of the same community label but has incident edges with a different community label. 

%brief overview of the proposed approach
\noindent{\underline{Brief Overview of the Proposed Approach}}: 
We propose a probabilistic model for detection of Holistic Community Outliers. We model the problem as a community detection task where the outliers are modeled as a separate community. A Hidden Markov Random Field (HMRF) is used to perform joint community analysis for both nodes and edges considering node data, edge data and the linkage structure. Thus, the HMRF contains both nodes and edges from the graph as vertices, referred to as a node-vertex and an edge-vertex respectively. Community labels for outliers are sampled from a uniform distribution while normal vertices could be sampled from Gaussian or a multinomial distribution. EM is used for the inference, and to compute the hidden community label $Z$ for every vertex. 

%summary of contributions
We make the following contributions in this paper:
\begin{itemize}
	\item We look at the community outlier detection problem from a holistic perspective by incorporating node data, edge data, and the linkage structure. Based on such a view, we propose the problem of detecting novel HCOutliers.
	\item We model the problem as a community analysis task over a HMRF and provide inference using an EM algorithm.
	\item Using several experiments on synthetic datasets and a DBLP dataset, we show the effectiveness of the proposed approach in identifying HCOutliers.
\end{itemize}

%organization of the paper.
Section~\ref{sec:related} gives a brief summary of related work. We describe our HMRF model in detail in Section~\ref{sec:approach} and present the inference in Section~\ref{sec:inference}. We discuss hyper-parameter settings and initialization details in Section~\ref{sec:discussions}. We provide details of experiments on synthetic datasets and the DBLP dataset in Section~\ref{sec:experiments} and conclude in Section~\ref{sec:conclusion}.

\section{Related Work}
\label{sec:related}

Outlier detection has a long history and~\cite{aggarwal13_book,chandola2009_csur,hodge2004_ai} provide extensive overviews of popular methods;~\cite{akoglu2015_dmkd,gupta2014outlier} focus on network outlier detection methods. Our work is most related to two main sub-areas of network outlier detection: community-based outlier detection and analysis of edge-attributed graphs, each of which is discussed below.

\subsection{Community-based Graph Outlier Detection}

Community-based outlier detection methods perform community analysis on graphs. Nodes that do not belong to any community are labeled as outliers. Community-based outlier detection has been studied both for static networks and dynamic networks. The notion of community outliers has been studied for static bipartite graphs using random walks in~\cite{sun2005_icdm}, for static homogeneous networks using probabilistic models in~\cite{gao2010_kdd}, and for static heterogeneous graphs using non-negative matrix factorization in~\cite{gupta2013community}. ~\cite{ranshous2015anomaly} summarizes different anomaly  detection methods in dynamic graphs such as decomposition based methods~\cite{akoglu2010event},~\cite{sun2006beyond}, distance based methods~\cite{DBLP:conf/icdm/AbelloED10}. Community based outlier detection method, for dynamic graphs, is proposed for a two-snapshots setting in~\cite{gupta12_kdd} and for a series of snapshots in~\cite{gupta12_pkdd}.

\subsection{Analysis of Edge-attributed Graphs}

Edge content indicates the type of relationship between the two nodes. However very little work exists on analysis of edge-attributed graphs. Qi et al.~\cite{qi2012_icde} proposed a edge-induced matrix factorization based method for finding communities in social graph using edge content.~\cite{beutel2013_www} proposed a method that considers edge data of User-Likes-Pages Facebook graph for temporal analysis to find the page-like pattern.~\cite{boden2012_kdd} proposed a method for mining coherent sub-graphs in multi-layer edges by exploiting edge content.~\cite{gupta2014top} proposed a method for finding top-$K$ interesting subgraphs matching a query template where interestingness is defined using edge weights. We propose a community-based outlier detection method for edge-attributed graphs which jointly exploits edge content along with node data and the linkage structure.
\section{Anomaly Detection using Node Data, Edge Data, and Linkage}
\label{sec:approach}

In this section, we develop the HMRF model which we use to perform joint community analysis for both nodes and edges considering node data, edge data and the linkage structure. Consider an undirected graph $G=\{V,E\}$, where $V$ is the set of vertices and $E$ is the set of edges. \\ %Table~\ref{tab:notations} lists the important notations used in this paper.

\noindent\underline{Node and Edge Data:} Let $S$ be the set of observed data.

\begin{itemize}
\item Let $S_i$ represent the observed data of node $v_i\in V$ which has $p$ attributes, $S_{i_1}$, $S_{i_2}$, $\cdots$, $S_{i_p}$.
\item Let $S_{ij}$ represent the observed data of edge $e_{ij}\in E$, i.e., edge data between nodes $v_i$ and $v_j$, and there are $q$ number of attributes, $S_{ij_1}$, $S_{ij_2}$, $\cdots$, $S_{ij_q}$.
\end{itemize}

\noindent\underline{HMRF Model:} The HMRF contains a vertex representing each node and each edge from $G$. We refer to the HMRF vertices representing nodes and edges as node-vertices and edge-vertices respectively. We use the index $i$ to refer to node-vertices in the HMRF, and the index $ij$ to refer to edge-vertices in the HMRF. We use the index $b$ to refer to all the vertices in the HMRF. Thus $b\in B=\{1, 2, \cdots, i, \cdots, |V|, (1,2), (1,3), \cdots, (1,|V|), (2,3), \cdots, (|V|-1,|V|)\}$. Note that $ij\in B$ if $i<j$. Thus, $|B|=|V|+|E|$.

\begin{itemize}
\item Let $X=\{X_{b}\}$ be a set of random variables. $X_{i}$ generates node data $S_i$, and $X_{ij}$ generates edge data $S_{ij}$.
\item Let $Z=\{Z_{b}\}$ be the set of hidden random variables. $Z_{i}$ indicates the community assignment of $v_{i}$. Suppose there are $K$ communities, then $Z_{i} \in \{0, 1, 2, \cdots, K\}$. If $Z_{i}=0$, then $v_{i}$ is an outlier. If $Z_{i}=k$, then $v_{i}$ belongs to the community $k$. Similarly, $Z_{ij}$ denotes the community assignment for edge $e_{ij}$.
\item Let $W$ denote the weights in the HMRF. $W_{v_i,v_j}$ is set to the weight of the edge $e_{ij}\in E$ and represent edge strength. $W_{v_i, v_j, e_{ij}}$ are weights for a triangle clique consisting of $v_i$, $v_j$ and $e_{ij}$. Thus, $|W|=4|E|$ (3 faces of the triangle formed by two node-vertices and an edge-vertex, and the whole triangle itself). 
\end{itemize}

$W_{v_i, e_{ij}}$ can be set in multiple ways. We set it to the inverse of number of neighbors of $v_i$ in $G$. Thus, for a co-authorship graph, this weight can be set to the inverse of the number of co-authors of author $v_i$ in the graph. For a Twitter users graph, this weight can be set to the inverse of the number of other users that the user $v_i$ is connected to. In a directed network sense, the weight can also be set to the ratio of the number of tweets from user $v_i$ to user $v_j$ to number of total tweets from user $v_i$. 

$W_{v_i, v_j, e_{ij}}$ should reflect some property of the triangle clique which cannot be captured using individual edges. For a co-authorship network, this could be set to ratio of number of papers where authors $v_i$ and $v_j$ are co-authors to the total number of papers on which either $v_i$ or $v_j$ are authors. Similarly, for a Twitter user network, this weight can be set to the ratio of tweets exchanged between users $v_i$ and $v_j$ to the total tweets posted by either $v_i$ or $v_j$.

$Z_b$'s are influenced by their neighbors, i.e., if two nodes are linked, it is likely that both belong to the same community. This does not hold for the outliers. Let $N_i$ denote the set of neighbors of node $v_i$. If $Z_i=0$, then $v_i$ is an outlier, and so the neighborhood set is empty. Also, each node-vertex is connected to the edge-vertex corresponding to the edges which are incident on the node in $G$.
\begin{align}
\small
N_{i} =
  \begin{cases}
    \{v_j|W_{v_i v_j}> 0, i \neq j, Z_{j}\neq 0 \} \cup\\
		     \{e_{xy}|W_{v_i e_{xy}}> 0, x=i\ or\ y=i, Z_{xy}\neq 0\}       & \quad Z_{i} \neq 0\\
    \phi  & \quad Z_{i}=0\\
  \end{cases}
\end{align}

Similarly, we also define neighbors for edge-vertices, $N_{ij}$, as follows. Note that for an edge-vertex, the neighbors are only the node-vertices on which the edge-vertex is incident except for the case when these node-vertices are outliers.

\begin{align}
\small
N_{ij} =
  \begin{cases}
    \{v_i|z_i\neq 0\}\cup\{v_j|z_j\neq 0\} & \quad Z_{ij} \neq 0\\
    \phi  & \quad Z_{ij}=0\\
  \end{cases}
\end{align}

\noindent\underline{Data Likelihood:} Let $\Theta$ = \{$\theta_{1},\theta_{2},..,\theta_{K}$\} be the set of parameters describing the normal communities. Thus, likelihood for node/edge data can be written as follows.

\begin{eqnarray}
P(X_{b}=S_{b}|Z_{b}=k)=P(X_{b}=S_{b}|\theta_{k})
\end{eqnarray}

We assume the outliers follow uniform distribution as we do not know beforehand which elements(node-vertex or edge-vertex) are anomalous or what they look like.

\begin{eqnarray}
P(X_{b}=S_{b}|Z_{b}=k)=\rho_{0}
\end{eqnarray}
\noindent where $\rho_{0}$ is a constant. 

%\begin{table*}
%\centering
%\caption{Important Notations}
%\label{tab:notations}
%\begin{tabular}{|l|l|} \hline
%Symbol & Definition\\ \hline
%%\texttt{$I=\{1, 2, \cdots, i, \cdots, M, 11, 12, \cdots, 1M, \cdots, MM\}$} & Indices for node data and edge data\\ \hline
%\texttt{$V=\{v_{1}, v_{2}, \cdots, v_{M}\}$}&Set of nodes\\ \hline
%\texttt{$E=\{e_{11}, e_{12}, \cdots, e_{1M}, \cdots, e_{M1}, \cdots, e_{MM}\}$}& Set of edges\\ \hline
%\texttt{$S_{i}=\{S_{i1}, S_{i2},\cdots, S_{ip}\}$}& Observed node attributes\\ \hline
%\texttt{$T_{ij}=\{T_{ij1}, T_{ij2}, \cdots, T_{ijq}\}$}&Observed edge attributes\\ \hline
%\texttt{$N=\{n_{1}, n_{2}, \cdots, n_{T}\}$}& Set of nouns (dictionary)\\ \hline
%\texttt{$W=[w_{ij}]$}& Matrix of size $(M^2 + M) \times (M^2 + M)$\\, representing edge strength between every pair of nodes in HMRF\\ \hline
%\texttt{$Z=\{Z_{1}, Z_{2}, \cdots, Z_{M}, Z_{11}, Z_{12}, \cdots, Z_{1M}, \cdots, Z_{MM}\}$}&Set of random variables for hidden labels of nodes in HMRF\\ \hline
%\texttt{$X=\{X_{1}, X_{2}, \cdots, X_{M}\}$}&Set of random variables for observed node data\\ \hline
%\texttt{$Y=\{Y_{11}, \cdots, Y_{1M}, \cdots, Y_{MM}\}$}& Set of random variables for observed edge data \\ \hline
%%\texttt{$X_{i}=\{X_{i_{age}}, X_{i_{location}}$\}}&Set of random variables for attributes age and location of node $v_i$\\ \hline
%\texttt{$N_i$}&Set of neighbor nodes of a node $i$ in HMRF\\ \hline
%\end{tabular}
%\end{
%table*}

\begin{figure}%
\centering
\subfloat{
\includegraphics[width=1\columnwidth]{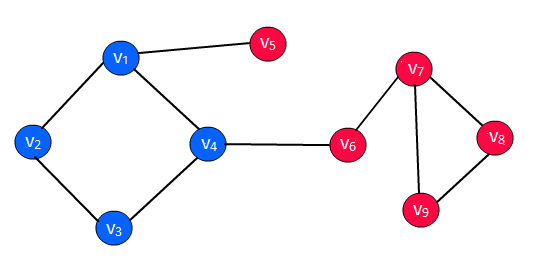}%
}
\caption{Original Graph G=(V,E)}%
\label{fig:img1_1}%

\subfloat{
\includegraphics[width=1\columnwidth]{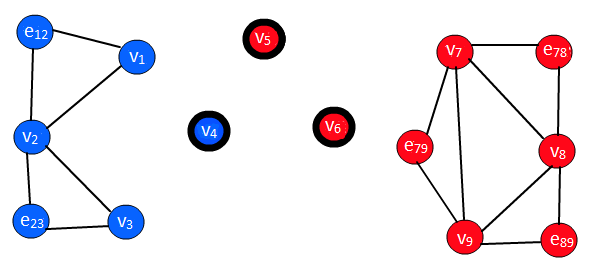}%
}
\caption{HMRF model of G indicating anomalous node $v_4$, $v_5$ and $v_6$}%
\label{fig:img1_2}%
\end{figure}

Figures~\ref{fig:img1_1} and ~\ref{fig:img1_2} illustrate a sample graph and its corresponding HMRF model. The graph has two communities, blue and red. The node $v_5$ has connection with the member of blue community as if it is part of blue community but its node attributes are similar to red community. Hence, in the HMRF model, it is not connected to any other node, indicating $v_5$ as an outlier. The edge between $v_4$ and $v_6$ indicate the cross-community collaboration and thus detected as interesting nodes.

\subsection{Cliques and Potentials in HMRF}
As the community label assignment follows an HMRF, we can define $P(Z) = \frac{1}{H_{1}} \exp(-U(Z))$ where $U(Z)$ is the potential function, defined as sum over all possible cliques. 

We consider two kinds of clique potentials: pairwise ($v_i$ to $v_j$, $v_i$ to $e_{ij}$) and triangular (between $v_i$, $v_j$ and $e_{ij}$). Then, the potential function can be written as follows.

\begin{align}
\small
\begin{split}
U(Z) = -\lambda_{1}\sum_{\substack {W_{v_{i}v_{j}}>0,\\Z_{i}\neq 0,\\Z_{j}\neq 0}}W_{v_{i}v_{j}}\delta(Z_{i}-Z_{j})\\
-\lambda_{2} \sum_{\substack {W_{v_{i}e_{ij}}>0,\\Z_{i}\neq 0,\\Z_{ij}\neq 0}}W_{v_{i}e_{ij}}\delta(Z_{i}-Z_{ij})\\
-\lambda_{2}\sum_{\substack {W_{v_{j}e_{ij}}>0,\\Z_{j}\neq 0,\\Z_{ij}\neq 0}}W_{v_{j}e_{ij}}\delta(Z_{j}-Z_{ij})\\
-\lambda_{3}\sum_{\substack{ W_{v_{i}v_{j}e_{ij}}>0,\\Z_{i}\neq 0,\\Z_{j}\neq 0,\\Z_{ij}\neq 0}}W_{v_{i}v_{j}e_{ij}}\psi(Z_{i},Z_{j},Z_{ij})
\end{split}
\label{eq:potential}
\end{align}

\noindent where $\lambda_{1}, \lambda_{2}, \lambda_{3}$ are constants, $W_{v_{i}v_{j}}>0$, $W_{v_{i}e_{ij}}>0$, $W_{v_{j}e_{ij}}>0$, $W_{v_{i}v_{j}e_{ij}}>0$, imply that there are links connecting $v_{i}$-$v_{j}$, $v_{i}$-$e_{ij}$, $v_{j}$-$e_{ij}$ and $Z_{i}$, $Z_{j}$, $Z_{ij}$ are non-zero. The $\delta$ function is defined as $\delta(x)=1$ if $x=0$ and $\delta(x)=0$ otherwise. Also, $\psi(Z_i, Z_j, Z_{ij})=1$ if $\delta(Z_{i}-Z_{j})+\delta(Z_{i}-Z_{ij})+\delta(Z_{j}-Z_{ij})\leq 1$, and 0 otherwise. Note that we define $\psi$ this way to capture the intuition that the clique potential should fire if at least two of the members in the clique share the same community label. Overall, the potential function suggests that, if members of the clique are normal objects, they are more likely to be in the same community when there exists a link connecting them in G, and the probability becomes higher if their link (weight) is stronger.

\subsection{Probability of Generating the Data}
Given the community label of a vertex ($Z_i$ or $Z_{ij}$) in the HMRF, the corresponding data ($X_i$ or $X_{ij}$) can be modeled as data generated from a distribution with parameters specific to the community. Community labels can be obtained using either Gaussian mixture model on the data if it is continuous, or as multinomial if the data is textual.\\

\noindent{\underline{Continuous Data:}} If the data is numeric and 1-dimensional, we need to learn the mean ($\mu_k$), and standard deviation ($\sigma_k$) for each community $k$ represented using a Gaussian. Given the parameters and the community label, the log likelihood can be written as follows.

\begin{eqnarray}
\ln P(X_{i}=S_{i}|Z_i=k)=-\frac{(S_i-\mu_k)^2}{2\sigma_k^2}-\ln \sigma_k -\ln \sqrt{2\pi}
\label{eq:app1}
\end{eqnarray}

In case of multi-dimensional data with $p$ dimensions, the log likelihood can be written as follows.

\begin{eqnarray}
\nonumber \ln P(X_{i}=S_{i}|Z_i=k) =-\frac{(S_i-\mu_k)^T \Sigma_k^{-1} (S_i-\mu_k)}{2}\\
-\frac{\ln \Sigma_k}{2} -\ln \sqrt{(2\pi)^p}
\end{eqnarray}

\noindent{\underline{Discrete Data:}} If the node (or edge) data can be represented as a document, we assume that the attributes (or words) are independent of each other, given the community label. 

\begin{eqnarray}
P(X_{i}=S_{i}|\theta_{k})=\prod_{c=1}^p P(X_{i_c}=S_{i_c}|\theta_{k})\\
P(X_{ij}=S_{ij}|\theta_{k})=\prod_{c=1}^q P(X_{ij_c}=S_{ij_c}|\theta_{k})
\label{eq:app2}
\end{eqnarray}

For text data, given a vocabulary of $T$ words, let $d_{bl}$ be the number of times the word $l$ appears in data related to vertex $b$. Then the parameters $\theta_k=\{\beta_{k1}, \cdots, \beta_{kT}\}$ denote the probability of a particular word belonging to community $k$. Given the parameters, the data likelihood of an object belonging to the $k^{th}$ community can be computed as follows.

\begin{eqnarray}
\ln P(X_{i_c}=S_{i_c}|Z_i=k)=\sum_{l=1}^T d_{il} \ln \beta_{kl}
\label{eq:app3}
\end{eqnarray}

\noindent where $\beta_{kl}$ is the probability of the word with index $l$ belonging to community $k$.
\section{Inferring $Z$ and Estimating $\Theta$}
\label{sec:inference}
The model parameters $\Theta$ and the set of hidden labels $Z$ are unknown as described in Section~\ref{sec:approach}. In this section, we describe methods to infer the hidden labels and also estimate the model parameters.

\subsection{Inferring Hidden Labels}
Assuming the model parameters $\Theta$ are known, we find the configuration $Z$ that will maximize the posterior distribution as follows.

\begin{eqnarray}
\hat{Z} = \arg\max_{\substack{Z}}P(X=S|Z)P(Z)
\label{eq:infer1}
\end{eqnarray}

To find such a configuration, we use the Iterated Conditional Modes (ICM) algorithm~\cite{besag1986statistical}. It is a greedy algorithm which guarantees convergence to a local optima by sequentially updating one $Z_b$ at a time assuming other $Z$'s (i.e., $Z_{B-{b}}$) as constant. At each step, the algorithm updates $Z_b$ given $X_b=S_b$ and the other labels such that the posterior $P(Z_b|X_b=S_b, Z_{B-{b}})$ is maximized. If $Z_{b}\neq 0$, applying Bayes rule, 

\begin{eqnarray}
P(Z_{b}|X_b=S_b,Z_{B-{b}})\propto P(X_{b}=S_{b}|Z)P(Z)
\label{eq:infer2}
\end{eqnarray}

In Eq.~\ref{eq:infer2}, $P(Z)$ can be computed using Eq.~\ref{eq:potential}. But rather than considering potential for the entire graph, Eq.~\ref{eq:infer2} needs potentials defined over only those cliques in which $Z_b$ is involved. Thus, we can write the following equation for node-vertices:

\begin{align}
\small
\begin{split}
P(Z_{i}|X_i=S_i,Z_{B-\{i\}}) & \propto P(X_{i}=S_{i}|Z) \times \\
& \exp(\lambda_{1}\sum_{\substack {v_{j}\in N_i}}W_{v_{i}v_{j}}\delta(Z_{i}-Z_{j})\\
& +\lambda_{2} \sum_{\substack {e_{xy}\in N_i}}W_{v_{i}e_{xy}}\delta(Z_{i}-Z_{xy})\\
&+\lambda_{3}\sum_{\substack{v_{j}\in N_i,\\e_{xy}\in N_i}}W_{v_{i}v_{j}e_{xy}}\psi(Z_{i},Z_{j},Z_{xy}))
\end{split}
\label{eq:infer3a}
\end{align}

Also, the corresponding equation for the edge-vertices can be written as follows.

\begin{align}
\small
\begin{split}
P(Z_{ij}|X_{ij}=S_{ij},Z_{B-\{(i,j)\}}) & \propto P(X_{ij}=S_{ij}|Z)\times\\
&\exp(\lambda_{2} \sum_{\substack {v_{i'}\in N_{ij}}}W_{v_{i'} e_{ij}}\delta(Z_{i'}-Z_{ij})\\
&+\lambda_{3} W_{v_{i}v_{j}e_{ij}}\psi(Z_{i},Z_{j},Z_{ij}))
\end{split}
\label{eq:infer3b}
\end{align}

Taking log of the posterior probability, we can transform the maximizing posterior problem to minimizing the conditional posterior energy function defined as shown in Eqs.~\ref{eq:infer4a} and~\ref{eq:infer4b} for node-vertices and edge-vertices respectively.

\begin{align}
\small
\begin{split}
U_i(k)&=-\ln P(X_{i}=S_{i}|Z_i=k)-\lambda_{1}\sum_{\substack {v_{j}\in N_i}}W_{v_{i}v_{j}}\delta(k-Z_{j})\\
&-\lambda_{2} \sum_{\substack {e_{xy}\in N_i}}W_{v_{i}e_{xy}}\delta(k-Z_{xy})\\
&-\lambda_{3}\sum_{\substack{v_{j}\in N_i,\\e_{xy}\in N_i}}W_{v_{i}v_{j}e_{xy}}\psi(k,Z_{j},Z_{xy})\\
\end{split}
\label{eq:infer4a}
\end{align}
\begin{align}
\small
\begin{split}
U_{ij}(k)&=-\ln P(X_{ij}=S_{ij}|Z_{ij}=k)\\
&-\lambda_{2} \sum_{\substack {v_{i'}\in N_{ij}}}W_{v_{i'} e_{ij}}\delta(Z_{i'}-k)-\lambda_{3} W_{v_{i}v_{j}e_{ij}}\psi(Z_{i},Z_{j},k)
\end{split}
\label{eq:infer4b}
\end{align}

If $Z_{b} = 0$, the vertex has no neighbors, and thus
\begin{align}
\begin{split}
P(Z_b|X_b=S_b,Z_{B-\{b\}})\  \propto\ P(X_{b}=S_{b}|Z_{b}=0)P(Z_{b}=0)
\end{split}
\end{align}

Hence, 
\begin{eqnarray}
U_b(0)=-\ln(\rho_0\pi_0)=a_0
\label{eq:infer5}
\end{eqnarray}

Finally, the label $Z_b$ for vertex $b$ in HMRF is set to $k\in \{0, 1, \cdots, K\}$ such that $U_b(k)$ is minimized. The predefined hyper-parameters, $\lambda_1,\lambda_2,\lambda_3$, represent the importance of the respective components of the graph structure. $a_0$ is the outlierness threshold. Algorithm~\ref{algo:inferLabels} first randomly initializes the labels for all vertices. At each step, labels are updated sequentially by minimizing $U_b(k)$.

\begin{algorithm}
%\scriptsize
\caption{Inferring Hidden Labels}
\textbf{Input:} Node/Edge data $S$, weights $W$, set of model parameters $\theta$, number of clusters $K$, hyper-parameters ($\lambda_1,\lambda_2,\lambda_3$), threshold $a_0$, initial assignment of labels $Z^{(1)}$\;\\
\textbf{Output:} Updated assignment of labels $Z$\;
\begin{algorithmic}[1]
\State Randomly set $Z^{(0)}$
\State $t\gets 1$
\While{$Z^{(t)}$ is not close enough to $Z^{(t-1)}$}
\State $t\gets t+1$
\For{$b=1;b\leq|B|;i++$}
\State Update $Z_b^{(t)}=k$ which minimizes $U_b(k)$ using Eqs.~\ref{eq:infer4a},~\ref{eq:infer4b} and ~\ref{eq:infer5}
\EndFor
\EndWhile\label{euclidendwhile}
\State \textbf{return} $Z^{(t)}$
\end{algorithmic}
\label{algo:inferLabels}
\end{algorithm}

\subsection{Estimating Parameters}
In this subsection, we discuss a method for estimating unknown $\theta$ from the data. $\theta$ describes the model that generates node data and edge data. Hence we seek to maximize the data likelihood $P(X=S|\theta)$ to obtain $\hat{\theta}$. However, since both the hidden labels and the parameters are unknown and they are inter-dependent, we use expectation-maximization (EM) algorithm (as shown in Algorithm~\ref{algo:em}) to solve the problem. We direct the reader to~\cite{gao2010_kdd} for details. 

The algorithm starts with an initial estimate of hidden labels $Z^{(1)}$. Given the current configuration of hidden labels, we can estimate model parameters as follows. For univariate numeric data, $\mu_k^{(t+1)}$ and $\sigma_k^{(t+1)}$ are estimated as follows:
\begin{eqnarray}
\mu_k^{(t+1)}=\frac{\sum_{b=1}^{|B|} P(Z_b=k|X_b=S_b, \Theta^{(t)}) S_b}{\sum_{b=1}^{|B|} P(Z_b=k|X_b=S_b, \Theta^{(t)})}\label{eq:muUpdate}\\
(\sigma_k^{(t+1)})^2=\frac{\sum_{b=1}^{|B|} P(Z_b=k|X_b=S_b, \Theta^{(t)}) (S_b-\mu_k)^2}{\sum_{b=1}^{|B|} P(Z_b=k|X_b=S_b, \Theta^{(t)})}
\label{eq:sigmaUpdate}
\end{eqnarray}

For text data, given a vocabulary of $T$ words, let $d_{bl}$ be the number of times the word $l$ appears in data related to vertex $b$. Then, given the current configuration of hidden labels, we can estimate parameters of the multinomial distribution $\beta_{kl}$ for $k\in \{1, \cdots, K\}$ and $l\in \{1,\cdots,T\}$ as follows:

\begin{eqnarray}
\beta_{kl}^{(t+1)}=\frac{\sum_{b=1}^{|B|} P(Z_b=k|X_b=S_b, \Theta^{(t)})d_{bl}}{\sum_{l=1}^T \sum_{b=1}^{|B|} P(Z_b=k|X_b=S_b, \Theta^{(t)}) d_{bl}}
\label{eq:betaUpdate}
\end{eqnarray}

Given the updated parameters, the new configuration of hidden labels can be estimated using Algorithm~\ref{algo:inferLabels} where $P(Z_b=k^*|X_b=S_b, \Theta^{(t)})=1$ if $k^*=\arg\min_k U_b(k)$, and 0 otherwise.

In summary, the Holistic Community Outlier detection algorithm shown in Algorithm~\ref{algo:em} works as follows. It begins with some initial label assignment of the vertices in the HMRF. In the M-step, the model parameters are estimated using the EM algorithm to maximize the data likelihood, based on the current label assignment. In the E-step, Algorithm~\ref{algo:inferLabels} is run to re-assign the labels to the objects by minimizing $U_b^{(k)}$ for each HMRF vertex sequentially. The E-step and M-step are repeated until convergence is achieved, and the outliers are the nodes that have 0 as the final estimated labels.

\begin{algorithm}
%\scriptsize
\caption{Holistic Community Outlier Detection}
\textbf{Input:} Node/Edge data $S$, weights $W$, set of model parameters $\theta$, number of clusters $K$, hyper-parameters ($\lambda_1, \lambda_2, \lambda_3$), threshold $a_0$, initial assignment of labels $Z^{(1)}$\;\\
\textbf{Output:} Set of Holistic Community Outliers\;
\begin{algorithmic}[1]
\State Randomly set $Z^{(0)}$
\State $t\gets 1$
\While{Z$^{(t)}$ is not close enough to Z$^{(t-1)}$}
\State \textbf{M-step:} Given $Z^{(t)}$, update parameters $\Theta^{(t+1)}$ according to Eqs.~\ref{eq:muUpdate},~\ref{eq:sigmaUpdate} or~\ref{eq:betaUpdate}.
\State \textbf{E-step:} Given $\Theta^{(t+1)}$, update the hidden labels as $Z^{(t+1)}$ using Algorithm~\ref{algo:inferLabels}.
\State $t\gets t+1$
\EndWhile
\State \textbf{return} the indices of outliers: $\{i:Z_b^{(t)}=0,b\in B\}$
\end{algorithmic}
\label{algo:em}
\end{algorithm}

\section{Discussions}
\label{sec:discussions}
In this section, we discuss how to set the hyperparameters and how to initialize the hidden labels.\\

\noindent{\underline{Setting Hyperparameters:}} The HCOutlier detection algorithm has the following hyperparameters: threshold $a_0$, clique importance variables, $\lambda_1,\lambda_2,\lambda_3$ and number of communities $K$. 

$\lambda_1,\lambda_2,\lambda_3$ indicate the importance of link between two nodes, the importance of link between edge-vertex and a node-vertex, and the importance of the triangle between the edge-vertex and the node-vertices of which the edge is incident respectively. If $\lambda$s are set to low values, the algorithm will consider only node information for finding the outliers. If their values are set too high, all connected nodes will have the same label, so an upper bound can be set for $\lambda_1+\lambda_2+\lambda_3$, such that a value above this bound will result into empty communities. High $\lambda_1$ will give more importance to the linkage in the graph. High $\lambda_2$ will give high importance to consistency between the node data and edge data in the graph. High $\lambda_3$ will give high importance to consistency between the edge data and both the incident nodes in the graph. 

The threshold $a_0$ can be replaced by another parameter (percentage of outliers $r$) as follows. In Algorithm~\ref{algo:inferLabels}, first calculate $\hat{Z}_i = \arg\min_k U_i(k) (k\neq 0)$ for each $i \in \{1, \cdots, |V|\}$ and then sort $U_i(\hat{Z}_i)$ in non-descending order. Finally, select top $r$ percentage as outliers.

$K$ represents the number of normal communities. For small value of $K$, algorithm will find the global outliers, and for large value of $K$ many local outliers will be detected because of many communities. An appropriate $K$ can be set using a variety of methods like Akaike Information Criterion (AIC), Bayesian Information Criterion (BIC), Minimum Description Length (MDL), etc.\\

\noindent{\underline{Initialization of labels:}} Instead of initializing $Z$ randomly, we initialize $Z$ values by clustering the nodes without considering outliers. To overcome local optima issues, we run the algorithm multiple times with different initialization, and choose the one with the largest data likelihood value.

\section{Experiments}
\label{sec:experiments}

Evaluation of outlier detection algorithms is difficult in general due to lack of ground truth. Hence, we perform
experiments on multiple synthetic datasets. We evaluate outlier detection accuracy of the proposed algorithm based on outliers injected in synthetic datasets. We evaluate the results on real datasets using case studies. We perform comprehensive analysis of objects to justify the top few outliers returned by the proposed algorithm. The code and the data
sets are publicly available\footnote{https://github.com/supriya-pandhre/HCODA}. All experiments are performed on a machine with Intel core i3-2330M processor running at 2.20GHz and 4GB RAM. 
%\url{https://www.dropbox.com/s/cp3iwi8h3u9t2k8/HCODA.zip?dl=0}

%\noindent\underline{Baseline: CODA~\cite{gao2010_kdd}} 

We compare our algorithm with Community Outlier Detection Algorithm (CODA)~\cite{gao2010_kdd}. CODA is also an algorithm for community-based outlier detection which performs community detection using node data and linkage; but ignores edge-data completely. 

\subsection{Synthetic Datasets}
We synthetically generate three graphs: Graph A ($|V|$=1000), Graph B ($|V|$=10000) and Graph C ($|V|$=100000). The node data is generated using 5 different Gaussian distributions and similar procedure is followed to generate edge data as well. Then for each of these graphs, different percentage of outliers are injected by randomly changing the data associated with 1\%, 5\% and 10\% of the nodes.

We generate a variety of synthetic datasets by varying the number of nodes in the graph, percentage of outliers injected in the graph and percentage of outliers to be extracted by the algorithm.  Table~\ref{tab:synAcc} shows the synthetic dataset results for the baseline (CODA) as well as the proposed algorithm (HCODA). We set the number of clusters ($K$) to 5 for these experiments (based on the generation process). We measure the performance of the two algorithms from two perspectives: firstly, accuracy of extraction of the injected outliers (OD Acc.) and secondly, how well it assigns the community label to all the nodes of the graph (CA Acc.). Note that each cell of the table corresponds to average over five runs of the algorithm. As the table shows, the proposed algorithm (HCODA) is 6.2\% more accurate than the baseline on average in terms of outlier detection accuracy. Similarly, HCODA is 2.2\% better than CODA on average. The performance is consistent across various graph sizes as well as across different degrees of outlierness in the graph. \\

\begin{table*}[pt]%
%\scriptsize
\caption{Synthetic Dataset Results on Graphs A, B and C (K=5; OD Acc.=Outlier Detection Accuracy, CA Acc.=Community Assignment Accuracy; CODA is Baseline, HCODA is the proposed algorithm)}
\label{tab:synAcc}
\scriptsize
\begin{tabular}{|c|p{0.90in}|p{0.70in}|p{0.70in}|p{0.70in}|p{0.70in}|p{0.70in}|p{1in}|}
%\begin{tabular}{|c|c|c|c|c|c|c|c|}
\hline
$|V|$&\% injected&\% extracted&CODA ~\cite{gao2010_kdd} OD Acc.&HCODA OD Acc.&CODA ~\cite{gao2010_kdd} CA Acc.&HCODA CA Acc.&Average runtime of HCODA (ms)\\
\hline
\hline
1000&1&1&1.000&1.000&0.714&0.713&898.6\\
\hline
1000&5&1&1.000&1.000&0.725&0.735&930\\
\hline
1000&5&5&0.760&0.800&0.739&0.756&899\\
\hline
1000&10&1&1.000&1.000&0.683&0.699&917.2\\
\hline
1000&10&5&0.960&0.980&0.720&0.738&913\\
\hline
1000&10&10&0.760&0.840&0.729&0.762&926\\
\hline
10000&1&1&0.690&0.730&0.770&0.784&9846.6\\
\hline
10000&5&1&0.970&1.000&0.740&0.755&10259\\
\hline
10000&5&5&0.746&0.780&0.758&0.777&10478.8\\
\hline
10000&10&1&1.000&1.000&0.705&0.718&10026.8\\
\hline
10000&10&5&0.958&1.580&0.741&0.757&9840.4\\
\hline
10000&10&10&0.795&0.814&0.759&0.777&9626.8\\
\hline
100000&1&1&0.722&0.735&0.773&0.787&120587\\
\hline
100000&5&1&0.995&0.996&0.747&0.762&113171.5\\
\hline
100000&5&5&0.751&0.789&0.764&0.783&113749.2\\
\hline
100000&10&1&0.997&0.999&0.698&0.714&104812.4\\
\hline
100000&10&5&0.946&0.974&0.733&0.751&103004\\
\hline
100000&10&10&0.784&0.808&0.750&0.770&100314\\
\hline
\end{tabular}
\end{table*}

\begin{figure}%
\includegraphics[width=\columnwidth]{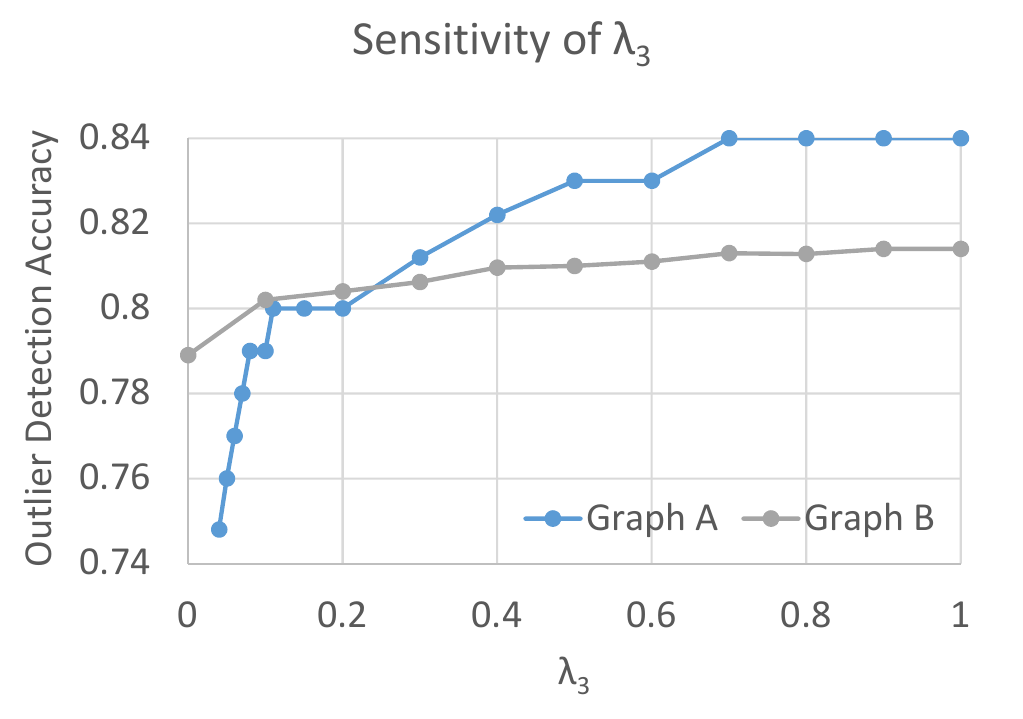}%
\caption{Sensitivity of $\lambda_3$}%
\label{fig:lambda3Sensitivity}%
\end{figure}

\begin{figure}%
\includegraphics[width=\columnwidth]{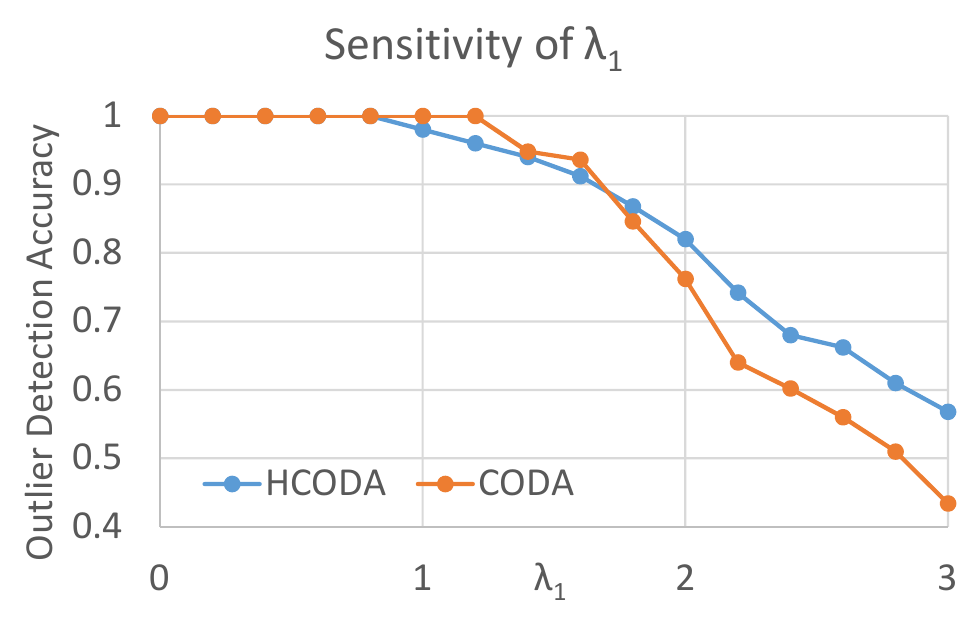}%
\caption{Sensitivity of $\lambda_1$ (Graph A)}%
\label{fig:lambda1Sensitivitya}%
\end{figure}

\begin{figure}%
\includegraphics[width=\columnwidth]{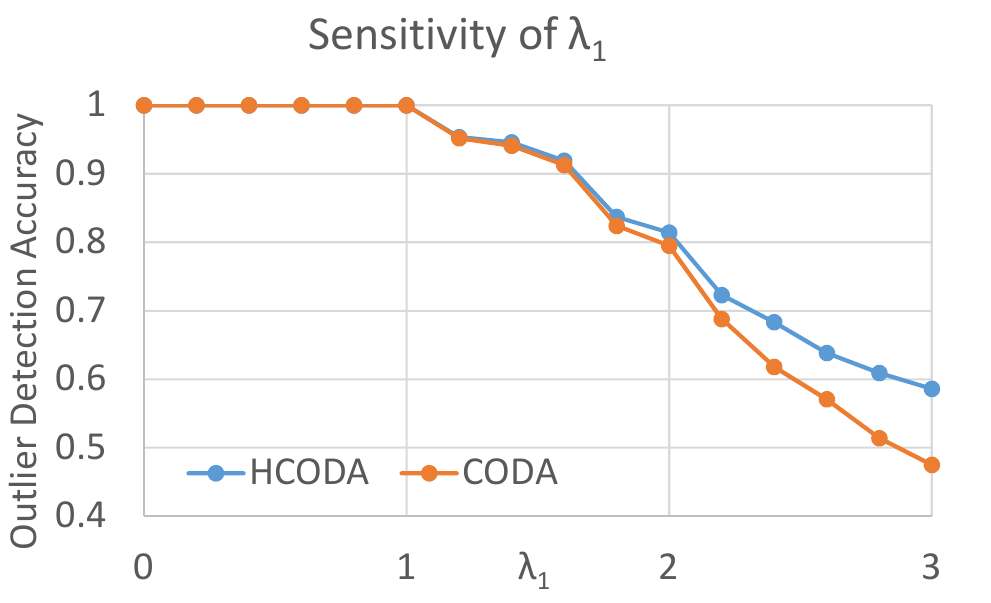}%
\caption{Sensitivity of $\lambda_1$ (Graph B)}%
\label{fig:lambda1Sensitivityb}%
\end{figure}

\noindent\underline{Hyperparameter Sensitivity:} To understand the impact of various hyperparameters ($\lambda$'s), we performed a few experiments. The results are shown in Figures~\ref{fig:lambda3Sensitivity},~\ref{fig:lambda1Sensitivitya} and~\ref{fig:lambda1Sensitivityb}. Figure~\ref{fig:lambda3Sensitivity} shows the sensitivity of the outlier detection accuracy with respect to variation of $\lambda_3$. We plot the curves for Graph A and B for the case of 10\% injected outliers and 10\% outliers to be extracted. Recall that $\lambda_3$ is the weight for the triangular clique. As can be seen from the figure, higher values of $\lambda_3$ are preferable but in general any value greater than 0.5 is reasonable.

Figures~\ref{fig:lambda1Sensitivitya} and~\ref{fig:lambda1Sensitivityb} show the sensitivity of the outlier detection accuracy with respect to variation of $\lambda_1$ for both the proposed algorithm (HCODA) and the baseline (CODA). Figures~\ref{fig:lambda1Sensitivitya} is the plot for Graph A while Figures~\ref{fig:lambda1Sensitivityb} is the one for Graph B. Both the plots are for the case of 10\% injected outliers and 1\% outliers to be extracted. Recall that $\lambda_1$ is the weight for the linkage in the original graph. As can be seen from the figure, lower values of $\lambda_1$ are preferable  both in the case of CODA as well as HCODA.

We noticed that the algorithm is not significantly sensitive to the parameter $\lambda_2$ and hence we do not show the corresponding plot.

\subsection{Four Area Dataset}
DBLP (\url{http://dblp.uni-trier.de/}) contains information about computer science journals and proceedings. Four Area is a subset of DBLP for the four areas of data mining (DM), databases (DB), information retrieval (IR) and machine learning (ML). It consists of papers from 20 conferences (five per area): KDD, PAKDD, ICDM, PKDD, SDM, ICDE, VLDB, SIGMOD, PODS, EDBT, SIGIR, WWW, ECIR, WSDM, IJCAI, AAAI, ICML, ECML, CVPR, CIKM. Thus, we consider the authors who have published papers in these conferences, and work with the co-authorship network (edge weight = co-authorship frequency). The edge data consisted of a vector of size four representing the count of papers co-authored in a particular research area. 

The graph contains a total of 42844 author nodes. Each author is associated with a vector of size 20 containing the count of papers published by an author in the twenty conferences. There is an edge between two authors if they co-authored a paper together and the count of such co-authored papers is the weight $W_{v_iv_j}$ between two nodes. 

The graph contains 118618 co-authorship edges. The weight of the link between the author-vertex and the co-authorship-edge-vertex in the HMRF (i.e., $W_{v_i e_{ij}}$) is set to the inverse of the number of co-authors of author $v_i$ in the graph.
$W_{v_i, v_j, e_{ij}}$ is set to ratio of number of papers where authors $v_i$ and $v_j$ are co-authors to the total number of papers on which either $v_i$ or $v_j$ are authors. We set the number of communities to $K$=4 (i.e., four normal communities and an outlier community). Also, we set the percentage of outlier parameter $r$ as 1\%. The average execution time of the lgorithm is 30.32 seconds.\\

\noindent\underline{Case Studies:} Result of algorithm show that it is able to identify interesting nodes(authors) and edges(co-authorship). We validate the result produced by algorithm by manually visiting the authors' homepages. For example, work by Sameer K. Antani is mainly in Clinical data standards and electronic medical records, but he published a paper with authors working in information retrieval i.e. node data of Sameer K. Antani is different from the fellow co-authors but the relationship is similar to what two authors from information retrieval should have and hence the algorithm has identified this node as interesting one. Similarly, Ivo Krka mainly focuses on software engineering and system modeling, however, he has published a paper on data mining and hence identified by the algorithm. 

The proposed algorithm also identified authors based on the edge information. We discuss a few cases now. Anindya Datta usually publishes in information systems, and Debra E. VanderMeer usually publishes work related to design science research, but their co-authored work is published on data management and very large databases topics. Sabyasachi Saha has co-authored papers with Sandip Sen on topics related to artificial intelligence and he has also published work on information and knowledge management with Pang-Ning Tan. The algorithm was able to identify him because of his work in multiple research areas.
\section{Conclusion}
\label{sec:conclusion}
In this paper, we introduced a method (HCODA) that detects novel Holistic Community Outlier graph nodes by taking into account the node data and edge data simultaneously to detect anomalies. We modeled the problem as a community detection task using a Hidden Random Markov Model, where outliers form a separate community. We used ICM and EM to infer the hidden community labels and model parameters iteratively. Experimental results show that the proposed algorithm consistently outperforms the baseline CODA method on synthetic data, and also identifies meaningful community outliers from the Four Area network data.

\bibliographystyle{plain}
\bibliography{references}
\end{document}